# Quantum key distribution over multicore fiber

**J. F. Dynes,[1] S. J. Kindness,[1,2] S. W.-B. Tam,[1] A. Plews,[1] A. W. Sharpe,[1] M. Lucamarini,[1] B. Fröhlich,[1] Z. L. Yuan,[1,*] R. V. Penty[2] and A. J. Shields[1]**

*[1]Toshiba Research Europe Ltd, Cambridge Research Laboratory,
208 Cambridge Science Park, Cambridge CB4 0GZ, UK*
*[2]Cambridge University Engineering Department, 9 JJ Thomson Ave, Cambridge, CB3 0FA, UK*
*\*zhiliang.yuan@crl.toshiba.co.uk*

**Abstract:** We present the first quantum key distribution (QKD) experiment over multicore fiber. With space division multiplexing, we demonstrate that weak QKD signals can coexist with classical data signals launched at full power in a 53 km 7-core fiber, while showing negligible degradation in performance. Based on a characterization of intercore crosstalk, we perform additional simulations highlighting that classical data bandwidths beyond 1Tb/s can be supported with high speed QKD on the same fiber.

**1. Introduction**

Quantum key distribution (QKD) is emerging as a powerful technique for secure distribution of encryption keys between remote parties [1-4]. It guarantees the security of the distributed keys using the fundamental constraints of quantum mechanics as well as providing the ability to detect an eavesdropper. Underpinned by many new developments, including the establishment of a quantified security level [2], QKD technology has matured considerably in the last decade. Furthermore the technology is making the transition from the laboratory to successful field trials [3-8].

As the bandwidth demands on data communication increases, multiplexing over different degrees of freedom in conventional single-core single-mode fiber (SSMF), including wavelength, phase, time and polarization multiplexing, is being utilized to circumvent the future information capacity crunch [9]. With help of advanced coding formats, these multiplexing techniques have allowed a channel capacity of up to 100 Tb/s per SSMF [10]. To prevent future saturation, space division multiplexing (SDM) was proposed to further increase the data bandwidth [11-14]. SDM can now be effectively realized by fabricating fibers with multiple single or few mode cores possessing low loss and crosstalk. These could replace standard SSMFs in telecoms systems to improve the bandwidth to the Pb/s regime [15-17].

Novel multicore fibers (MCF's) are also an excellent opportunity for QKD. Previously, operation of QKD systems usually required dedicated SSMFs, which are not only expensive but might even be unavailable. Multiplexing QKD into data-populated SSMF has been explored, but it constrains either the communication distance or the launched data laser power and maximum bandwidth [8, 18-25]. These constraints are a consequence of signal contamination to the quantum channel by the Raman photons generated by the strong data signals in the same fiber. Future MCF deployment could allocate one core for quantum signals only, therefore permitting simultaneous transmission of QKD and data signals through the same fiber without severe restriction on the data launch power or bandwidth. This opportunity has not been explored to date, however.

In this paper we report the first QKD experiment over MCF. After a careful characterization of crosstalk of a 7-core MCF of 53-km length, we run QKD continuously for a duration of 24 hours in the presence of spatially multiplexed bi-directional 10 Gb/s data signals launched at 0 dBm power. We compare the MCF results with a dual SSMF setup and find no performance penalty associated with the use of MCF. Moreover, a simulation based on the measured parameters shows that MCF can support a simultaneous transmission of QKD and Tb/s data in the same fiber.

## 2. Multicore fiber characterization

The MCF selected for the experiment is 53 km in length, and has a cladding diameter of 185 µm and 7 cores with a core-to-core distance of 47 µm, see Fig. 1(a). Each core supports a single optical mode, and has similar transmission characteristics as a standard telecom SSMF. For example, the central core of the MCF is measured to have an attenuation of 0.23 dB/km at 1550 nm, which is only slightly higher than the SSMF which has a typical loss rate of 0.20 dB/km. It is noteworthy that the MCF's diameter is only a factor of 1.5 larger than that of a standard SSMF. At the same time, the MCF displays a 7-fold increase in achievable bandwidth due to the increased spatial capacity. This highlights MCF's potential to increase the communication bandwidth when incorporating into current telecoms systems [15-17].

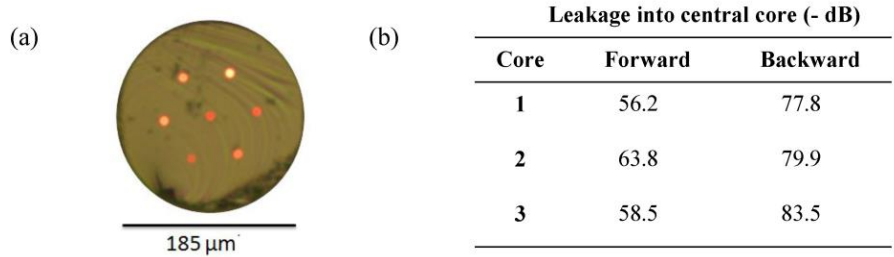

Fig. 1. (a) Optical microscope images showing cross-sections of the multicore fiber (MCF); The MCF was manufactured by OFS Labs. (b) Power leakage into the central core measured in decibels from 3 outer cores. A 1550 nm laser is used in this measurement.

Optical crosstalk between MCF cores must be carefully characterized, because of its possible impact on the quantum channel which operates in the single photon regime. We consider two types of crosstalk. First, photons transmitted in one core can transfer into adjacent cores through evanescent field coupling. We term this type as *leakage*, and the leaked photons have the same wavelength as the launched signal. The second type is *Raman crosstalk*, scattered from neighboring cores where strong data signals are present. To characterize these two types of crosstalk, we assign the central core of the MCF as the quantum channel and the outer cores for classical signals. This configuration allows all classical signals to impact the quantum channel equally, and thus is the most suitable to evaluate the MCF's applicability to QKD. A laser power of 0 dBm (1 mW) is launched into one of the outer cores at a time, and we then measure the optical power exiting from the central one at either the launch end (i.e.: backward scattered signal) or the receive end of the MCF fiber (i.e.: forward scattered).

We measure the leakage signal from different outer cores with results summarized in Fig. 1(b). Here, the leakage value is calculated as the ratio of the optical power exiting from the central core to the launch power into the outer one. The forward and backward leakage signals are around 60 and 80 dB weaker than the launched signal, respectively. The latter is lower because Rayleigh backscattering is required in addition to intercore evanescent field coupling. At the given data launch power of 0 dBm, the forward leakage level shown in Fig. 1(b) corresponds to a noise intensity on the order of 10 photons/ns, which is about 100 times stronger than the received quantum signal expected for a weak-pulse QKD system. This level of contamination therefore necessitates applying noise rejection techniques. Thankfully, the leakage signals have well-defined wavelength, which is the same as that of the data laser, so they can be effectively removed spectrally by a standard DWDM filter in the quantum receiver. It has been previously shown [21] that even 0 dBm data signal within the same fiber core can be effectively filtered.

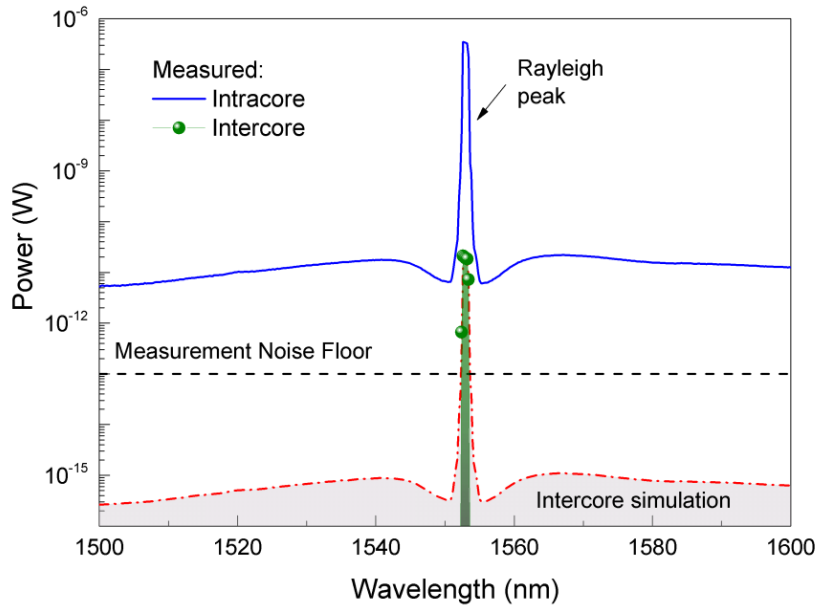

Fig. 2. Spectra of backscattered signals into the same outer core (intracore, blue line) or the central core (intercore, green dots) in the presence of a 0 dBm power of a 1552-nm laser launched into an outer core. The dashed line shows the noise floor of the optical spectrum analyzer used in the measurement. The dash-dotted line shows the simulated Raman crosstalk.

We attempt to measure the Raman crosstalk in the backward configuration only. This is because we expect backward scattering to be stronger than forwards scattering as the backscattered signal is attenuated less by the fiber. We use an optical spectrum analyzer (OSA) to spectrally resolve the signal scattered into the central core. However, this intercore Raman signal is below the noise floor of the OSA and too weak to be measured. As shown in Fig. 2, we resolve only a sharp peak (green dots), caused by the backward leakage, at the same wavelength as the launch laser. To deduce the Raman crosstalk, we instead measure the backward Raman signal exiting from the same outer core. As shown in Fig. 2, the intracore spectrum is essentially identical to that of a SSMF [18]. The Raman signal distributes over a wide spectrum exceeding 100 nm, and its peak intensity is about 40 dB lower than the Rayleigh peak. To quantify the Raman crosstalk we assume that it has the same spectral shape as for intercore scattering, however, with a 40 dB lower intensity as suggested by the difference of inter- and intracore Rayleigh peak. The Raman crosstalk is below 1 fW over the entire spectral range. This level of noise is negligible relative to the quantum channel power and is expected not to affect the operation of QKD.

## 3. QKD/data spatial multiplexing setup

Figure 3 shows our experimental setup. The 53-km, 7-core MCF spool forms the optical channels between two communicating parties, Alice and Bob. At each end of the fiber, all transmitters or receivers are terminated with SSMF's. Optical coupling between SSMF's and MCF is realized through two 1×7 fanouts featuring a coupling loss in the range of 1.0 – 4.6dB.

We use a QKD system operating with a clock rate of 1 GHz and using the T12 decoy-state protocol [26]. The photons are detected with InGaAs avalanche photodiodes operating in Geiger mode and with self-differencing [27] electronics. While crucial for noise rejection in SSMF multiplexing [18], the temporal filtering capability offered by the detectors is less important in the present spatial multiplexing setup. Optical crosstalk from other channels (cores) is much less than in the SSMF case since space division gives >40dB isolation. The QKD system utilises a fully automatic FPGA control for stabilising the photon count rate and quantum bit error rate (QBER) through feedback control of the polarization controller, fiber phase delay and system delay. Error correction is accomplished using Cascade [28] and a number theoretic transform approach to execute efficient matrix multiplication is utilized for privacy amplification.

The QKD system requires a quantum channel and auxiliary classical channels for clock synchronization and key reconciliation. We assign two cores of the MCF for these signals: the central core is to carry the weak quantum signal while one of the outer cores for the QKD auxiliary classical signals. The spare 5 outer cores are free to use for intense classical data communication signals. The quantum signal is transmitted on the DWDM grid with a wavelength of 1547.72 nm while all classical signals are transmitted at other wavelengths. This arrangement allows rejection of the intercore leakage into the quantum channel using a standard DWDM filter of 0.4 nm passband located in the quantum receiver. A pair of 10 Gb/s data channels, with 0 dBm power each launched into separate cores from opposite directions, are assigned at the wavelength of 1552.72 nm.

The quantum channel has a total loss of 14.1 dB, consisting of 12.4 dB due to the MCF, 1.1 dB loss at the optical fanout and 0.6 dB at the DWDM filter.

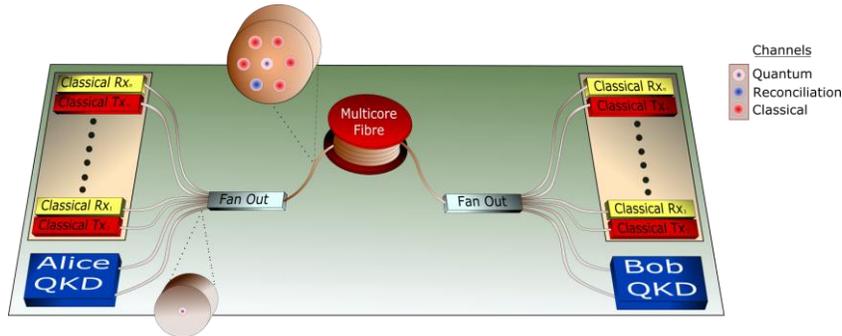

Fig. 3. QKD over MCF experimental setup.

## 4. Results and discussion

We operate the QKD system over the central core of the MCF while bi-directional 10Gb/s classical signals are transmitted simultaneously through two of the six outer cores. A bit-error tester is used to monitor the data communication. It typically does not record a single error during the experiments. Figure 4 plots the secure key rate and quantum bit error rate (QBER) over a 24 hour period. These values are obtained through real-time data-processing, including error correction and privacy amplification, on a sifted block size of $1 \times 10^8$ bits. The block size is sufficiently large to achieve 85% of the asymptotic secure key rate [26]. Each data block is collected over a session time of about 36 seconds with a sifted bit rate of ~2.7 Mb/s. Over the entire measurement, the QBER displays a very small fluctuation around its average value of 3.36% with a standard deviation of 0.54%. The secure bit rate fluctuates more visibly, because of its sensitivity to the QBER with an average secure key rate of 605±18 kb/s. This result represents the first QKD operation over MCF, and also the first demonstration of spatial mode multiplexing of data and QKD over the same fiber.

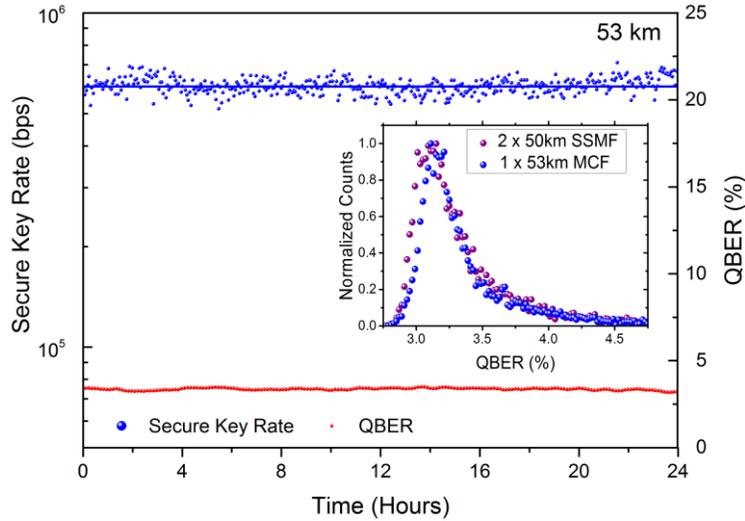

Fig. 4. QKD results over 53 km of MCF in the presence of 10 Gb/s bi directional classical traffic launched at 0 dBm in outer cores. The solid blue line indicates the average value of the secure bit rate. Inset: Comparison of the histograms of instantaneous QBERs obtained with MCF and standard SSMF's.

We compare the above MCF result with a control experiment over standard SSMF's. We use two 50-km SSMF spools to replace the MCF. Quantum signals are transmitted through one spool, while all classical signals are transmitted through the other. The physical separation of these signals ensures zero crosstalk. Extra attenuation is added to the quantum channel to be consistent with the MCF experiment. All other experimental parameters remain identical to the MCF case. QKD over the SSMF's yields a secure bit rate of 619 kb/s and a QBER of 3.36% averaged over a duration of four hours. Within statistical error, there is no significant impairment in performance using the MCF compared to two SSMFs. This demonstrates that neither the MCF, nor the intense 10G data signals on the same MCF cause a performance penalty, except for the higher attenuation of the MCF which may be improved with better quality fiber and coupling in the future. The inset of Fig. 4 compares the distributions of instantaneous QBERs recorded with a ~0.5 second sampling time, to detect difference in the disturbance to the quantum signal by different fibers. Nearly identical distributions suggest that the MCF has a very similar optical characteristics to standard telecom SSMF.

We simulate the impact of higher power data signals on the QKD performance assuming equal-power bidirectional data launched into the separate outer cores. As the Raman coefficient is wavelength dependent, we choose a worst-case scenario in which both forward and backward data channels produce the maximum Raman scattering to the quantum channel. The scattering coefficient is obtained from the intracore scattering measurement corrected by the amplitude difference in Rayleigh peaks (Fig. 2). Parameters for the QKD system are taken directly from experimental values, including clock frequency, channel attenuation, detector efficiency and dark count rate.

Figure 5 shows the simulation results. In the absence of data channels, our simulation gives a theoretical secure key rate of 627 kb/s and a QBER of 3.36%. These values do not degrade noticeably for an increasing combined power of data channels of up to 100 mW, or 20 mW per core. Accounting for the optical crosstalk into the quantum channel, this power level is well below the point at which non-linear effects are expected in DWDM transport systems [19]. Moreover, they are in excellent agreement with measured values in both MCF (2× 1 mW) and SSMF (crosstalk free) experiments. Further increasing the data launch power causes a gradual increase of QBER and hence a reduction in the secure key rate. Nevertheless, the simulation still predicts the co-existence of QKD with more than 1 W of data laser power in each direction. This highlights the potential of the MCF spatial multiplexing technique to support QKD along with high bandwidth data transport. DWDM systems containing 128 channels are now commercially available [29] and if a full-power 10Gb/s per channel system was deployed in each of the 5 spare outer cores, this would correspond to a combined power of only 320 mW in each direction and a total of 6.4 Tb/s of bidirectional classical information simultaneously transmitted with a high rate QKD channel.

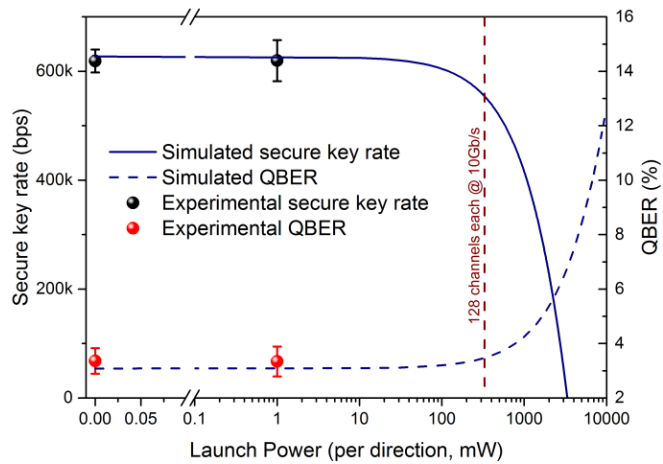

Fig. 5. Simulation of QKD performance over the 53 km MCF in the presence of bidirectional data communication, using the worst-case scenario of Raman crosstalk into the quantum channel. Also shown is the total launch power of 1 mW, 128× 10 Gb/s channels (brown dashed line).

## 5. Conclusion

In this paper, QKD has been successfully demonstrated through MCF for the first time. Our experiment demonstrates that standard 10 Gb/s bi-directional traffic can be spatially multiplexed onto the same MCF without detrimental effect on the performance of the QKD. Secure key rates averaging at 605 kb/s are successfully transmitted over 53 km of MCF continuously for a 24 hour period. Furthermore, our calculation suggests that QKD can potentially coexist with Tb/s bidirectional data traffic on the same fiber.

**Acknowledgment**

S.J.K. acknowledges support from the EPSRC CDT in Photonic Systems Development.